\documentclass[aps,prl,amsmath,twocolumn]{revtex4}
\usepackage{CJK}
\usepackage{amsmath}
\usepackage{enumerate}
\usepackage{amssymb}
\usepackage{amsfonts}
\usepackage{mathrsfs}
\usepackage{latexsym}
\usepackage{bm}
\usepackage{amscd}
\usepackage{graphicx}
\usepackage{epsfig}
\usepackage{hhline,multirow}
\usepackage{dcolumn}
\usepackage{url}
\usepackage[colorlinks=true,linkcolor=blue]{hyperref}
\usepackage{color}
\usepackage{indentfirst}
\usepackage{subfigure}
\usepackage{psfrag}
\usepackage{slashed}
\usepackage{float}
\usepackage{setspace}

\begin{document}

\title{Sivers distribution functions of sea quark in proton with chiral Lagrangian}
\author{Fangcheng He$^{1,2}$}
\author{P. Wang$^{1,3}$}
\affiliation{$^1$Institute of High Energy Physics, CAS, P. O. Box
918(4), Beijing 100049, China}
\affiliation{$^2$School of Physics, University of Chinese Academy of Sciences, Beijing 100049, China}
\affiliation{$^3$Theoretical Physics Center for Science Facilities, CAS,
		 Beijing 100049, China}

\begin{abstract}

We propose a mechanism for the Sivers distribution function in proton with chiral Lagrangian.
By introducing the gauge link of the vector meson, the transverse momentum dependent distribution of a pion in the nucleon
is re-defined which is locally $SU(2)_V$ invariant as the Lagrangian. The eikonal propagator is generated from the gauge link
and this scenario is proved to be equivalent to the final state interaction.
By combining the calculated splitting function and the valence $\bar{q}$ distribution in $\pi$ from the recent fit,
the sea quark Sivers function in proton is obtained. We find reasonable
numerical results for the first momentum $x\Delta^Nf_{\bar{q}}^{(1)}(x)$ without any
fine tuning of the free parameters.

\end{abstract}


\maketitle

In the recent decades, the transverse partonic structure of hadrons has been the subject of a lot of
theoretical and experimental investigations. The so called transverse momentum dependent (TMD) parton distributions
are of great interest since they offer insight in the three-dimensional structure of hadrons in terms 
of the QCD degrees of freedom \cite{Pisano:2015kaa}. 
At leading twist there are totally eight TMD parton distributions. Among them, two distributions, i.e., 
Boer-Mulders (BM) and Sivers distributions are time-reversal odd \cite{Boer:1997nt}. Compared with the BM distributions,
more data of Sivers distributions were extracted from semi-inclusive deep inelastic scattering (SIDIS) 
collected by HERMES and COMPASS collaborations \cite{Airapetian:2009ae,Bradamante:2017vuv}.
Sivers function describes the asymmetric distribution of unpolarized quarks
in a transversely polarized parent hadron. It is very essential to explain the
single-spin asymmetries (SSAs) in SIDIS which have been observed experimentally 
since a long time \cite{Sivers:1989cc,Bravar:1999rq,Airapetian:2001eg}. 

Theoretically, it is very difficult to calculate parton distribution functions (PDFs) from the first principle
due to the nonperturbative behavior of QCD.
Since PDFs are defined in Minkowski space, originally, it is also impossible to simulate on the Euclidean lattice.
Though the quasi-PDFs are proposed to be calculated on lattice based on the large momentum effective theory (LaMET)  \cite{Ji:2013dva},
the simulation of PDFs on Lattice is still in the early stage. For the Sivers distribution function,
most calculations are based on the phenomenological quark models, such as spectator model 
\cite{Brodsky:2002cx,Boer:2002ju,Bacchetta:2003rz,Lu:2004au,Goeke:2006ef,Lu:2006kt,Bacchetta:2008af,Maji:2017wwd}, MIT bag model \cite{Yuan:2003wk},
constituent quark model \cite{Courtoy:2008mj,Pasquini:2010af}, etc.
In these model calculations, the gluon field is introduced as the gauge link. The T-odd parton distributions
are zero without this gauge link because of the time reversal invariance. Dynamically, T-odd PDFs emerge from 
the gauge link structure of the parton correlation functions which describe the initial/final state interactions
\cite{Belitsky:2002sm,Collins:2002kn}. 

In deep inelastic scattering (DIS) process, the calculations of meson cloud effects were performed by Sullivan, where
the nucleon is composed of mesons (pions, kaons) and a bare baryon \cite{Sullivan:1971kd}. 
It is well known that effective field theory (EFT) is a very good and systematic method to study hadron physics.
There are a lot of applications of EFT on hadron spectrum, form factors and hadron-hadron interaction. 
In particular, for the parton distributions, it can be obtained from the convolution form, where 
the splitting function can be derived with chiral Lagrangian \cite{Burkardt:2012hk,Salamu:2018cny}. 
Without fine tuning, the obtained PDFs as well
as the integrated moments are in reasonable agreement with the experimental data \cite{Salamu:2014pka,Wang:2016eoq}.
However, there is no such kind of calculation for the T-odd TMD PDFs 
with EFT. The reason is that on the one hand, if we use the same approach,
the splitting function is zero for the Sivers distributions. One the other hand, the colored gluon 
field introduced from the gauge link is not consistent with the framework of EFT which is formulated 
in terms of hadronic degrees of freedom.

Therefore, in this paper, we will provide a mechanism to generate the T-odd TMD PDFs with chiral Lagrangian.
The bilocal operator constructed for the splitting function is invariant under the flavor $SU(2)$ symmetry instead of 
color gauge symmetry. With this approach, we will calculate the sea quark Sivers distribution functions in proton 
which have not been estimated theoretically even in the quark models. Sea quark Sivers functions are important to
explain azimuthal asymmetries for $\pi^{\pm}$ and $K^{\pm}$ production off a proton target in SIDIS 
and the asymmetrical cross sections for vector boson in polarized Drell-Yan process \cite{Adamczyk:2015gyk,Huang:2015vpy}.
It is also crucial to test the sign change of Sivers functions between SIDIS and Drell-Yan process.
Though the sea quark Sivers functions have been extracted from the experiments 
\cite{Bacchetta:2011gx,Anselmino:2016uie,Martin:2017yms,Boglione:2018dqd}, 
the theoretical explanation is still lacking.
The calculation here is for the Sivers functions of sea quark in proton and 
it is straightforward to generally apply it to any T-odd distributions. 

\begin{figure}[tbp]
\begin{center}
\includegraphics[scale=0.6]{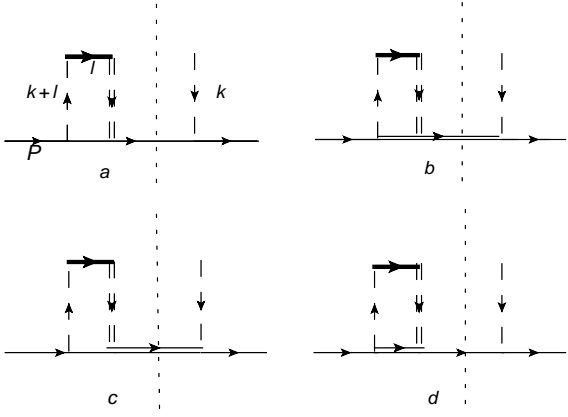}
\caption{The Sivers distribution functions of pseudoscalar mesons in the nucleon. 
The solid, dashed, double dashed, and double solid lines are for the octet baryons, pseudoscalar mesons, 
vector mesons and decuplet baryons respectively.
The thick solid line is the eikonal propagator and the dotted line means the on-shell cut.}
\label{diagrams}
\end{center}
\end{figure}

For the quark flavor $q$, according to the Trento convention, the unpolarized and
Sivers distributions $f_1(x,\vec{k}_\perp)$ and $f_{1T}(x,\vec{k}_\perp)$ are defined as \cite{Bacchetta:2004jz}
\begin{eqnarray}\label{pdf} 
f_1^q(x,\vec{k}_\perp)&+&\frac{\epsilon^{ji}k_\perp^iS_\perp^j}{m_p}f_{1T}^{q}(x,\vec{k}_\perp)
=\frac12\int\!\frac{d\xi^-d^2\vec{\xi}_\perp}{(2\pi)^3}  \nonumber \\
&\times& e^{-ixP^+\xi^-+i\vec{k}_\perp\cdot\vec{\xi}_\perp} \langle P,\vec{S}_\perp\mid\mathcal{O}^q\mid P,\vec{S}_\perp\rangle,
\end{eqnarray}
where $\vec{S}_\perp$ is the transverse spin of proton. The gauge invariant bilocal operator $\mathcal{O}^q$ is defined as \cite{Ji:2002aa}
\begin{eqnarray}\label{inv}
\mathcal{O}^q=\bar{q}(\xi^-,\vec{\xi}_\perp)\mathcal{L}_{\xi_\perp}^{\dag}(\infty,\xi^-)\gamma^+
\mathcal{L}_0(\infty,0)q(0,0),
\end{eqnarray}
where $\mathcal{L}$ is the path-ordered light-cone color gauge link expressed as
\begin{eqnarray}
\mathcal{L}_{\xi_\perp}^{\dag}(\infty,\xi^-)=P\,e^{-ig_c\int_{\xi^-}^{\infty}A^+(z^-,\vec{\xi}_\perp)dz^-}.
\end{eqnarray}
Similar as the quark distribution, for the pion distribution, i.e., the splitting function, the operator can be defined 
from the light-cone bilocal meson operator as
\begin{equation}\label{nm}
\mathcal{O}^{\pi^+} = i\big[\pi^-(y^-,\vec{y}_\perp)\partial^+\pi^+(0)-\partial^+ \pi^-(y^-,\vec{y}_\perp)\pi^+(0)\big].
\end{equation}
This kind of operator based on hadronic degrees of freedom has been applied for the calculation of pion distributions 
in the EFT \cite{Chen:2001nb,Salamu:2014pka}. However, the above operator gives no contribution to the T-odd Sivers function.
Therefore, we need to construct a bilocal operator for the meson fields which has the time reversal asymmetry. 
In Refs.~\cite{Bando:1984ej,Bando:1987br,Tanabashi:1995nz}, vector meson $V_\mu$ is introduced as a dynamical gauge boson to guarantee the 
local $SU(2)_V$ hidden symmetry. The matrix of $V_\mu$ is written as
\begin{eqnarray}
V^\mu=\vec{\rho}\,^\mu\cdot\vec{\tau}=\left(
\begin{array}{lcr}
\frac1{\sqrt{2}}\rho^0 & \rho^+  \\
~~\rho^- & -\frac1{\sqrt{2}}\rho^0  \\
\end{array}
\right)^\mu.
\end{eqnarray}
The Lagrangian for the meson fields can be written as
\begin{eqnarray}\label{lag}
 \mathcal{L}=f_\pi^2Tr[\alpha_{\mu}\alpha^{\mu}],
\end{eqnarray}
where $\alpha_{\mu}$ is defined as
\begin{eqnarray}
\alpha_{\mu}=\frac{1}{2i}(D_\mu\xi_L\cdot\xi_L^\dag-D_\mu\xi_R\cdot\xi_R^\dag).
\end{eqnarray}
The covariant derivatives are expressed as  
\begin{eqnarray}
D_\mu\xi_{L/R}=\partial_\mu\xi_{L/R}+igV_\mu\xi_{L/R}+i\xi_{L/R}L_\mu/R_\mu,
\end{eqnarray}
where $L_\mu$ and $R_\mu$ are the external fields. The coupling constant $g=g_{\rho\pi\pi}/\sqrt{2}$  
and $g_{\rho\pi\pi}$ is related to the vector meson mass $M_\rho$ through
the Kawarabayashi-Suzuki-Fayyazuddin-Riazuddin  
relation $M_\rho^2=2g_{\rho\pi\pi}^2 f_\pi^2$ \cite{Kawarabayashi:1966kd,Riazuddin:1966sw}. 
$f_\pi=92.1$ MeV is the pion decay constant \cite{Tanabashi:2018oca}.
In the chiral Lagrangian, $\xi_L^\dag=\xi_R=\xi=\,$exp(i$\pi/f_\pi)$ with $\pi = \vec{\pi}\cdot\vec{\tau}/\sqrt{2}$.
When matching quark currents to hadron level, $L_\mu$ and $R_\mu$ are expressed as $L_\mu=R_\mu=\tau_qv_\mu$,
where $\tau_q = \text{diag} (\delta_{qu},\delta_{qd})$ are diagnal $2 \times 2$ quark flavour matrices.
$v_\mu$ is the external vector field. From Eq.~(\ref{lag}),
we can get the local current for a given quark flavor \cite{Salamu:2018cny}. 
To get the bilocal operator at hadron level, the nonlocal action is written as 
\begin{eqnarray}\label{nonlocal}
\mathcal{S}=\int dx\,dy f_\pi^2Tr[\alpha_\mu(y)W(y,x)\alpha^{\mu}(x) W^\dag(y,x)],
\end{eqnarray}
where the gauge link function $W(y,x)$ is introduced to guarantee the nonlocal Lagrangian is
locally $SU(2)_V$ invariant as the local one. $W(y,x)$ is defined as
\begin{equation}
W(y,x)=Pe^{I(x,y)}=Pe^{-ig\int_x^y d z_\nu V^\nu(z)}.
\end{equation}
At the leading order of $g$, the current that couples to the external field $v_\mu$ can be obtained from Eq.~(\ref{nonlocal}) as
\begin{eqnarray}\label{current}
& &\hspace*{-0.7cm}\mathcal{J}_{q/\pi}^{\mu}=2i\,\text{Tr}\{[\partial^\mu\pi(y)\pi(x)-\pi(y)\partial^\mu\pi(x)]\tau_q  \nonumber \\
&&\hspace*{-1.2cm}+\,\partial^\mu\pi(y)[I(x,y),\pi(x)\tau_q]+\partial^\mu\pi(x)[I(x,y),\tau_q\pi(y)]\}.
\end{eqnarray} 
For example, the current for the $\bar{d}$ quark in $\pi^+$ is written as
\begin{eqnarray}\label{current2}
\mathcal{J}_{\bar{d}/\pi^+}^{\mu}&=&i[\pi^-(y)\partial^\mu\pi^+(x)-\partial^\mu\pi^-(y)\pi^+(x)\big] \nonumber\\
 &\times&\big[1-i\sqrt{2}g\int_x^y d z_\nu\rho^{0\nu}(z)\big].
\end{eqnarray}
From the above equation, one can see the quark current that couples to the external vector field
is expressed in hadronic degrees of freedom. With this matching, we can calculate the quark distribution function
in proton using the convolution form. The splitting function or the pion distribution
in the convolution form is obtained from the pion 
operator $\mathcal{O}^{\pi^+}_{tot}$. It can be
separated into two terms as 
\begin{eqnarray}\label{op}
\mathcal{O}^{\pi^+}_{tot}=\mathcal{O}^{\pi^+}+\mathcal{O}^{\pi^+}_{Sivers},
\end{eqnarray}
where 
\begin{eqnarray}\label{gl}
& &\hspace*{-0.8cm}\mathcal{O}^{\pi^+}_{Sivers}=\big[\pi^-(y^-,\vec{y}_\perp)\partial^+\pi^+(0)-\partial^+
\pi^-(y^-,\vec{y}_\perp)\pi^+(0)\big]  \nonumber \\
& &\hspace*{-0.6cm}\times \sqrt{2}g\{\int_0^\infty dz^-\rho^{0+}(z^-,0)+\int_\infty^{y^-} dz^-\rho^{0+}(z^-,\vec{y}_\perp)\}.
\end{eqnarray}
The first term in the above equation $\mathcal{O}^{\pi^+}$ is the ordinary bilocal pion operator defined in Eq.~(\ref{nm}). 
For the T-even distributions, this term is dominant and vector meson
contribution from the second term can be ignored. However, for the Sivers distribution function, $\mathcal{O}^{\pi^+}$
gives no contribution and $\mathcal{O}^{\pi^+}_{Sivers}$ is crucial to get the nonzero value.
With the above operator on the hadronic degrees of freedom, we can get the Sivers distribution function of a pion 
in proton $f_{1T}^{\pi/p}(z,\vec{k}_{\perp\pi})$ as
\begin{eqnarray}\label{Siverpi}
\hspace*{-2cm}& &\frac{\epsilon^{ji}k_{\perp\pi}^iS_\perp^j}{m_p}f_{1T}^{\pi/p}(z,\vec{k}_{\perp\pi})
=\frac12\int\,\frac{dy^-d^2\vec{y}_\perp}{(2\pi)^3}    \nonumber \\ 
\hspace*{-0.7cm}&\times&e^{-i(zP^+y^--\vec{y}_\perp\cdot\vec{k}_{\perp\pi})} 
\langle P,\vec{S}_\perp\mid \mathcal{O}^\pi_{Sivers} \mid P,\vec{S}_\perp\rangle.
\end{eqnarray}
It can be calculated with the chiral Lagrangian and
the leading loop diagrams are plotted in Fig.~1, where the solid, dashed, double dashed and double solid lines are 
for octet baryons, pseudoscalar meson, vector meson and decuplet baryons, respectively. 
The thick solid line is the eikonal propagator and the dotted line means the on-shell cut. 
The effective $\pi NN$ and $\pi N\Delta$ interaction can be 
written as $(g_A/\sqrt{2}f_\pi)\bar{N}\slashed{\partial}\gamma_5\vec{\pi}\cdot\vec{\tau}N$ and 
$(6g_A/5f_\pi)\bar{N}[g_{\mu\nu}+z\gamma_\mu\gamma_\nu]\partial^\nu\vec{\pi}\cdot\vec{I}\Delta^\mu$ \cite{Scherer:2002tk}.
$g_A=1.26$ is the axial charge. $z$ is the off-shell parameter and our results are independent of $z$ because 
the intermediate decuplet is on-shell. There are several ways of incorporating vector
mesons into chiral Lagrangians \cite{Schechter:1986vs,Yamawaki:1986zz,Meissner:1986tc}. In this paper,
the Lagrangians for the $\rho NN$, $\rho\Delta\Delta$ and $\rho N\Delta$ 
interactions are obtained from Refs.~\cite{Jenkins:1991ts,Geng:2009ys,Jones:1972ky} 
by substituting $eQA_\mu$ with $gV_\mu$. They are expressed as
\begin{eqnarray}\label{rD}
& &\hspace*{-0.8cm}\mathcal{L}_{\rho NN}=-g\bar{N}(\gamma^\mu-\kappa_N\frac{\sigma^{\mu\nu}\partial_\nu}{2m_N})\vec{\rho}_\mu\cdot\vec{\tau}N, \nonumber \\
& &\hspace*{-0.8cm} \mathcal{L}_{\rho \Delta\Delta}=-g\bar{\Delta}_\alpha(\gamma^{\alpha\beta\mu}+g^{\alpha\beta}\kappa_\Delta\frac{\sigma^{\mu\nu}\partial_\nu}{2m_\Delta})\vec{\rho}_\mu\cdot\vec{\Sigma}\Delta_\beta, \nonumber \\
& &\hspace*{-0.8cm}  \mathcal{L}_{\rho N\Delta}=-i\frac{gG^M_{N\Delta}}{2m_N}\bar{N}\gamma^\mu\gamma^5(\partial_\mu\vec{\rho}_\nu-\partial_\nu\vec{\rho}_\mu)\cdot\vec{I}\Delta^\nu+h.c.,
  \end{eqnarray}
where $1+\kappa_\Delta=\frac{3}{5}(1+\kappa_N)$, $G^M_{N\Delta}=\frac{6\sqrt{2}}{5}(1+\kappa_N)$ according to 
the quark model \cite{Brown:1975di} and the value of $\kappa_N$ is $6.1\pm0.2$ \cite{Mergell:1995bf}.  
$\vec{\Sigma}$ and $\vec{I}$ are the isospin 3/2 and isospin transition matrices 
\cite{Haidenbauer:2017sws}.
For the intermediate octet baryons, the contribution to $f_{1T}^{\pi/p}(z,\vec{k}_{\perp\pi})$ is written as
 \begin{eqnarray}\label{T}
& &\hspace*{-0.5cm}\frac{\epsilon^{ji}k_{\perp\pi}^iS_\perp^j}{m_p}f_{1T}^{\pi/p}(z,\vec{k}_{\perp\pi})=\frac{ig^2 g_A^2}{4f_\pi^2}
\int\!\frac{d^4k}{(2\pi)^4}\int\!\frac{d^4l}{(2\pi)^4}\bar{U}(P,\vec{S}_\perp)   \nonumber\\
& &\hspace*{-0.5cm}\slashed{k}\gamma^5S_{\text{on}}(P-k)
V_\mu(l)S(P-k-l)\gamma^5(\slashed{k}+\slashed{l})U(P,\vec{S}_\perp)
\nonumber\\
&\times &\frac{(2k^++l^+)}{(l^++i\epsilon)}
S_\pi(k)S_\rho^{\mu+}(l)S_\pi(k+l)   \nonumber\\
&\times&\delta(k^+-zP^+)
\delta^2(\vec{k}_\perp-\vec{k}_{\perp\pi})+h.c.,
\end{eqnarray}
where $V_\mu(l)$ is the vertex of the interaction between nucleon and $\rho$ meson expressed as 
$V_\mu(l)=\gamma_\mu+i\kappa_N\frac{\sigma_{\mu\nu}l^\nu}{2m_N}$. $S$, $S_\pi$ and $S_\rho^{\mu+}$ are
the propagators of nucleon, $\pi$ and $\rho$, respectively. $S_{\text{on}}$ is the on-shell nucleon propagator
expressed as $S_{\text{on}}(k) = 2\pi(\slashed{k} +m_N)\delta(k^2-m_N^2)$. The imaginary part of the eikonal
propagator $1/(l^+ + i\epsilon)$ gives the real Sivers distribution function of a pion in the nucleon.
The expressions for the other diagrams with decuplet intermediate states are similar but more complicated. 
\begin{figure}[tbp]
\begin{center}
\includegraphics[scale=0.55]{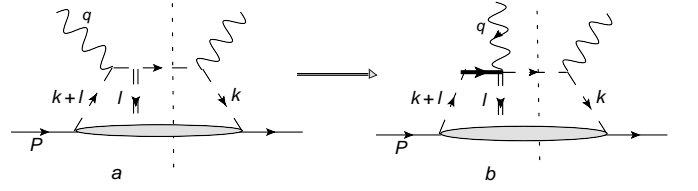}
\caption{Final-state interaction in Sullivan process with collinear approximation. 
The dashed, double dashed and wave lines are for the pseudoscalar mesons, vector mesons and photons respectively. 
The gray bubble represents octet and decuplet baryons.}
\label{diagrams}
\end{center}
\end{figure}

The nonzero Sivers distribution function can also be explained from the final state interaction (FSI).
Our approach can be applied in a non-perturbative QCD regime, where the final state interaction 
is described by the pion-baryon interaction.  
The left diagram in Fig.~2 denotes the FSI in Sullivan process,
where the momentum $k$ and $l$ are collinear with proton momentum $P$ in collinear approximation.
The `+' component of the momentum is much larger than the other components. As a result, the vector meson projects 
into `+' direction at leading order. In other words, the leading part of the momentum of the pseudoscalar meson 
after the photon scattering is the `$-$' component. This is also consistent with the analysis of parton distributions 
in Ref.~\cite{Belitsky:2002sm}. 
Therefore, the $\rho \pi\pi$ vertex and pseudoscalar propagator turn into the eikonal propagator approximately as
\begin{eqnarray}
\frac{(2k+2q+l)^-}{(k+q+l)^2-M^2}\approx \frac{(2k+2q+l)^-}{(2k+2q+l)l+i\epsilon}\approx\frac{1}{l^++i\epsilon}.
\end{eqnarray}
Accordingly, the diagram Fig.~2a can be changed into Fig.~2b, which means the final-state interaction effect has been 
absorbed into the distribution functions of pion in the nucleon. 
As a result the Sivers distribution function of a pion in the nucleon can be extracted 
and it is consistent with the gauge link approach in Fig.~1. 
In the calculation with spectator model based on the quark-gluon interaction, 
similar diagram as Fig.~2b is plotted to show that the effect of final-state interaction can be
absorbed into the distribution functions of the target nucleon \cite{Boer:2002ju}.

With the above splitting function, the Sivers distribution function of sea quark $\bar{q}$ in proton can be 
obtained by the convolution form, where the sea quark distributions can be expressed in terms of 
splitting function and quark distribution in pion \cite{Salamu:2014pka,Wang:2016eoq}.
For the TMD distributions, the convolution form is similar. For example, the anti-down quark Sivers function 
in proton can be expressed as
\onecolumngrid 
\begin{eqnarray}\label{dbt}
\frac{\epsilon^{ji}k_\perp^iS_\perp^j}{m_p}f_{1T}^{\bar{d}/p}(x,\vec{k}_\perp)&=&\frac{ig^2g_A^2}{4f_\pi^2}\int_0^1 \frac{dz}{z}\theta(z-x)\int\,\frac{d^4ld^4k_\pi}{(2\pi)^8}\delta(zP^+-k_\pi^+)\bar{U}(P,\vec{S}_\perp)\slashed{k}_\pi\gamma^5S_{\text{on}}(P-k_\pi)V_\mu(l)    \nonumber \\
&\times&S(P-k_\pi-l)\gamma^5(\slashed{k}_\pi+\slashed{l})U(P,\vec{S}_\perp)\frac{2k_\pi^+}{(l^++i\epsilon)} S_\pi(k_\pi)S_\rho^{\mu+}(l)S_\pi(k_\pi+l)   \nonumber \\
&\times&\frac{-1}{2} \int \frac{d^4l_1}{(2\pi)^3}Tr[\gamma^+(-\slashed{l_1}+m_q)\Gamma(k_\pi, l_1)(\slashed{k}_\pi-\slashed{l_1}+m_q)
\Gamma(k_\pi,l_1)(-\slashed{l_1}+m_q)]    \nonumber \\
&\times& \delta((k_\pi-l_1)^2-m_q^2)\delta(l_1^+-\frac{x}{z}k_\pi^+)\delta^2(\vec{l_1}_\perp-\vec{k}_{1\perp}-\frac{x}{z}\vec{k}_{\perp\pi}),
\end{eqnarray}
\twocolumngrid 
where $\Gamma(k_\pi,l_1)$ is the quark-meson coupling vertex.
The first two rows on the right hand side of the above equation correspond to the $\pi$ Sivers function written in Eq.~(\ref{T}), while the last two rows are for the anti-down quark distribution in $\pi$ defined as 
\begin{eqnarray}\label{val}
&& f^{\bar{d}/{\pi}}_{1v}(y,\vec{k}_{1\perp})= -\frac12\int\!\frac{d\xi^-d^2\vec{\xi}_\perp}{(2\pi)^3} \nonumber \\
&\times& e^{iyk_\pi^+\xi^--i(\vec{k}_{1\perp}+y\vec{k}_{\pi\perp})\cdot\vec{\xi}_\perp} \langle \pi^+\mid\mathcal{O}^d\mid \pi^+\rangle,
\end{eqnarray}
where $\mathcal{O}^d=\bar{d}(\xi^-,\vec{\xi}_\perp)\gamma^+d(0)$.
Therefore, Eq.~(\ref{dbt}) can be expressed by the convolution form as
\begin{eqnarray}\label{TMD}
& &\hspace*{-2cm}k_\perp^if^{\bar{q}/p}_{1T}(x,\vec{k}_\perp)=\int\,d^2\vec{k}_{\perp\pi}k^i_{\perp\pi}   \nonumber\\
& &\hspace*{-1.5cm}\int_x^1\frac{dz}{z}f^{\bar{q}/\pi}_{1v}(\frac{x}{z},
\vec{k}_\perp-\frac{x}{z}\vec{k}_{\perp\pi})f^{\pi/p}_{1T}(z,\vec{k}_{\perp\pi}),
\end{eqnarray}
where $f^{\bar{q}/\pi}_{1v}(\frac{x}{z},\vec{k}_\perp-\frac{x}{z}\vec{k}_{\perp\pi})$ is the 
quark TMD distribution in pion with the intrinsic transverse momentum $\vec{k}_\perp-\frac{x}{z}\vec{k}_{\perp\pi}$.
The first moment of the Sivers distribution function is defined as \cite{Anselmino:2016uie}
\begin{eqnarray}\label{qs}
& &\hspace*{-1cm}\Delta^Nf^{(1)}_{\bar{q}}(x)=\int\,d^2\vec{k}_\perp\frac{-k_\perp^2}{2m_p^2}f^{\bar{q}/p}_{1T}(x,\vec{k}_\perp)  \nonumber\\
& &\hspace*{-1cm}=\frac{1}{2m_p^2}\int_1^x d(\frac{x}{z})f_{1v}^{\bar{q}/\pi}(\frac{x}{z})\int\,d^2\vec{k}_{\perp\pi}
\vec{k}^2_{\perp\pi}f^{\pi/p}_{1T}(z,\vec{k}_{\perp\pi}),
\end{eqnarray}
where $f^{\bar{q}/\pi}_{1v}(x)$ is the quark distribution in $\pi$ and it can be
obtained from the recent fit at $Q=0.63$ GeV \cite{Aicher:2010cb}.

In the numerical calculation, the dipole regulator $\tilde F_j(k)$ ($j=\pi,\rho$) is applied to deal with the 
ultraviolet divergence \cite{Salamu:2018cny,Machleidt:1989tm}
\begin{eqnarray}
\tilde F_j(k)=\left(\frac{M_j^2-\Lambda_j^2}{k^2-\Lambda_j^2}\right)^2.
\end{eqnarray}
For pion case, in Ref.~\cite{Barry:2018ort}, the monopole regulator is chosen and
the corresponding $\Lambda_\pi$ is 0.52 GeV. Here because we include the decuplet intermediate state,
the monopole regulator is not sufficient to get rid of the UV divergence in Fig.1b. 
From the previous calculation of electromagnetic form factors, 
strange form factors and asymmetry of sea quark distributions of proton, reasonable $\Lambda_\pi$ 
in the dipole regulator is around 1 GeV \cite{He:2018eyz,Yu}. 
For $\rho$ meson, the parameter $\Lambda_\rho$ was chosen to be 1.85 GeV in Ref.~\cite{Machleidt:1989tm}.
Therefore, we present the results for the range  0.8 GeV $ \leq \Lambda_\pi \leq $ 1.2 GeV
and 1.6 GeV $ \leq \Lambda_\rho \leq $ 2.0 GeV. We should mention that
with the regulator, there is no power counting included in our method.

\begin{figure}[tbp]
\begin{center}
\includegraphics[scale=0.9]{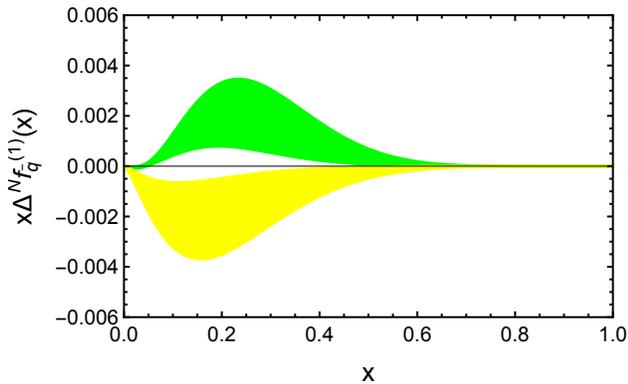}
\caption{The first momentum of sea quark Sivers distribution functions versus $x$ at $Q=0.63$ GeV. The green and yellow bands are the 
results of $x\Delta^Nf_{\bar{d}}^{(1)}(x)$ and $x\Delta^Nf_{\bar{u}}^{(1)}(x)$ 
with $0.8$ GeV $\leq \Lambda_\pi \leq 1.2$ GeV and $1.6$ GeV $\leq \Lambda_\rho \leq 2.0$ GeV. }
\label{diagrams}
\end{center}
\end{figure}

\begin{figure}[tbp]
\begin{center}
\includegraphics[scale=0.9]{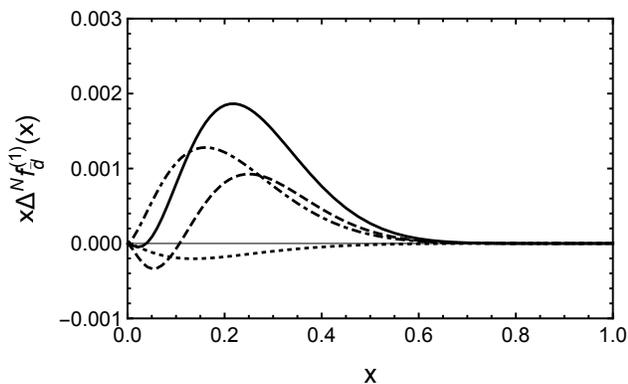}
\caption{Contributions to $x\Delta^Nf_{\bar{d}}^{(1)}(x)$ from different intermediate states with 
$\Lambda_\pi=1$ GeV, $\Lambda_\rho=1.85$ GeV and $\kappa_N=6.1$. The dashed, dotted and dot-dashed lines are for 
the contributions from intermediate octet, decuplet and octet-decuplet transition, 
respectively. The solid line is for the total result.}
\label{diagrams}
\end{center}
\end{figure}

The first moment of the Sivers distribution functions of $\bar{d}$ and $\bar{u}$ is plotted in Fig.~3. 
The green and yellow bands are for $x\Delta^Nf_{\bar{d}}^{(1)}(x)$ and $x\Delta^Nf_{\bar{u}}^{(1)}(x)$, respectively. 
For $\bar{d}$ in proton, the first moment is positive. The maximum value of $x\Delta^Nf_{\bar{d}}^{(1)}(x)$ is
$0.0008 - 0.0035$ at $x$ around 0.2. It then decreases with increasing $x$ and when $x$ is larger than 0.6, 
$x\Delta^Nf_{\bar{d}}^{(1)}(x)$ will tend to be zero. 
For $\bar{u}$ in proton, $x\Delta^Nf_{\bar{u}}^{(1)}(x)$ is always negative. The maximum absolute value
is about $0.0007 - 0.0037$ at $x$ around 0.15. Similar as for $x\Delta^Nf_{\bar{d}}^{(1)}(x)$, when $x$ is larger than 0.6, 
$x\Delta^Nf_{\bar{u}}^{(1)}(x)$ will also approach to zero. 
As many phenomenological extractions, the value of sea quark Sivers function is very small \cite{Bacchetta:2011gx, Martin:2017yms}. 

Our result is consistent with the prediction in the large $N_C$ limit where the absolute values of the Sivers distribution functions
of $\bar{d}$ and $\bar{u}$ are the same while their signs are opposite \cite{Pobylitsa:2003ty}.
In Ref.~\cite{Bacchetta:2011gx}, where the data are extracted at $Q = 1$ GeV,
the central value of $x\Delta^Nf_{\bar{d}}^{(1)}(x)$ is negative, while $x\Delta^Nf_{\bar{u}}^{(1)}(x)$ is positive.
Considering the sign difference in the definition of the first moment between our Eq.~(\ref{qs}) and Eq.~(4) 
in Ref.~\cite{Bacchetta:2011gx}, the two results are consistent with each other.
Compared with the results in Ref.~\cite{Anselmino:2016uie}, where the central values of extracted $x\Delta^Nf_{\bar{u}}^{(1)}(x)$ and 
$x\Delta^Nf_{\bar{d}}^{(1)}(x)$ 
are both negative and the absolute value of $x\Delta^Nf_{\bar{d}}^{(1)}(x)$ is much larger than $x\Delta^Nf_{\bar{u}}^{(1)}(x)$,
our $x\Delta^Nf_{\bar{d}}^{(1)}(x)$ has similar magnitudes but with opposite sign. 
For $x\Delta^Nf_{\bar{u}}^{(1)}(x)$, the sign is the same as their best fit but our magnitude is larger. 
Hopefully, these differences can be checked by the further theoretical and experimental analysis.

For $x\Delta^Nf_{\bar{u}}^{(1)}(x)$, only one diagram Fig.~1b gives contribution. However, for $x\Delta^Nf_{\bar{d}}^{(1)}(x)$,
all the four diagrams in Fig.~1 give contribution. To see the separate contribution clearly, 
we plot the contribution to $x\Delta^Nf_{\bar{d}}^{(1)}(x)$ from different intermediate states. 
The dashed, dotted and dot-dashed lines are for the contributions from intermediate octet, decuplet and octet-decuplet transition, 
respectively. The solid line is the total result.
From the figure, we can see that for $x\Delta^Nf_{\bar{d}}^{(1)}(x)$, 
the contributions from intermediate octet and octet-decuplet transition are dominant.
The contribution of  the octet-decuplet transition gives large positive value to $x\Delta^Nf_{\bar{d}}^{(1)}(x)$. 
For the contribution from the octet intermediate state, the sign is 
$x$ dependent. It is negative at small $x$ and when $x>0.1$, the sign changes to be positive. 
The contribution to $x\Delta^Nf_{\bar{d}}^{(1)}(x)$ from the decuplet intermediate state is 
very small. The decuplet intermediate state gives negative contribution to both
$x\Delta^Nf_{\bar{d}}^{(1)}(x)$ and $x\Delta^Nf_{\bar{u}}^{(1)}(x)$. However, the
contribution to $x\Delta^Nf_{\bar{d}}^{(1)}(x)$ is 9 times smaller than that to $x\Delta^Nf_{\bar{u}}^{(1)}(x)$
due to the smaller value of the coupling constants for $\pi^+$ case than for $\pi^-$ case.

Our calculation with the chiral Lagrangian is valid at low energy scale. The result
is supposed to hold up to the scale of $\rho$ meson mass or 1 GeV ($4\pi f_\pi$). This is why the input scale of 
the pion PDF is chosen to be at 0.63 GeV.
The scale evolution of the Sivers function as well as the first momentum $\Delta^Nf^{(1)}_{\bar{q}}(x)$ is 
discussed in \cite{Aybat:2011ge,Anselmino:2012aa,Kang:2012em}.
With the scale increasing, the maximum of $x\Delta^Nf_{\bar{d}}^{(1)}(x)$ and $x\Delta^Nf_{\bar{u}}^{(1)}(x)$ will become
smaller due to the effect of diagonal terms in the twist-3 evolution kernel \cite{Boglione:2018dqd}. 

In summary, we proposed a mechanism for the study of Sivers distribution function with chiral Lagrangian. 
The vector meson is introduced for the $SU(2)_V$ hidden symmetry. 
The bilocal $\pi$ operator is re-defined with the gauge link of the vector meson which is locally $SU(2)$
invariant. The eikonal propagator generated from the flavor gauge link is crucial to obtain a nonzero Sivers distribution
function. The gauge link approach is also proved to be consistent with
the final state interaction in the collinear approximation. With the convolution form, which combining the
splitting function calculated from the bilocal $\pi$ operator and the valence quark distribution
in pion, the Sivers distribution functions of $\bar{u}$ and $\bar{d}$ are obtained. Numerical results show that 
the absolute values of $x\Delta^Nf_{\bar{u}}^{(1)}(x)$ and $x\Delta^Nf_{\bar{d}}^{(1)}(x)$ 
are close to each other, while their signs are opposite. For $x\Delta^Nf_{\bar{d}}^{(1)}(x)$, 
the contributions from intermediate octet state and octet-decuplet transition are dominant. The decuplet intermediate state
gives negligible contribution. For $x\Delta^Nf_{\bar{u}}^{(1)}(x)$ the only contribution 
comes from the decuplet intermediated state and it is 9 times larger than the corresponding
contribution for $x\Delta^Nf_{\bar{d}}^{(1)}(x)$. Without any fine
tuning of the parameters, our results are consistent with the
prediction obtained in the large NC limit, and are also comparable
with the recent phenomenological extractions from fitting the
experimental data. This is the first theoretical estimation 
on $x\Delta^Nf_{\bar{u}}^{(1)}(x)$ and $x\Delta^Nf_{\bar{d}}^{(1)}(x)$ within the framework of chiral Lagrangian.
Our predictions can be checked by the future theoretical and experimental analysis.

\section*{Acknowledgments}

This work is supported by the National Natural Sciences Foundations of China under the grant No. 11475186,
the Sino-German CRC 110 ``Symmetries and the Emergence of Structure in QCD" project by NSFC under the grant No.11621131001, 
and the Key Research Program of Frontier Sciences, CAS under grant No. Y7292610K1.


\begin{thebibliography}{10}

\bibitem{Pisano:2015kaa}
S.~Pisano
\href{http://dx.doi.org/10.1051/epjconf/20158502033}{{\em EPJ Web Conf.}
  {\bfseries 85} (2015) 02033}.

\bibitem{Boer:1997nt}
D.~Boer and P.~J. Mulders
  \href{http://dx.doi.org/10.1103/PhysRevD.57.5780}{{\em Phys. Rev.} {\bfseries
  D57} (1998) 5780--5786},
\href{http://arxiv.org/abs/hep-ph/9711485}{{\ttfamily arXiv:hep-ph/9711485
  [hep-ph]}}.

\bibitem{Airapetian:2009ae}
{\bfseries HERMES} Collaboration, A.~Airapetian {\em et~al.}
  \href{http://dx.doi.org/10.1103/PhysRevLett.103.152002}{{\em Phys. Rev.
  Lett.} {\bfseries 103} (2009) 152002},
\href{http://arxiv.org/abs/0906.3918}{{\ttfamily arXiv:0906.3918 [hep-ex]}}.

\bibitem{Bradamante:2017vuv}
{\bfseries COMPASS} Collaboration, F.~Bradamante
  \href{http://dx.doi.org/10.1088/1742-6596/938/1/012004}{{\em J. Phys. Conf.
  Ser.} {\bfseries 938} no.~1, (2017) 012004},
\href{http://arxiv.org/abs/1711.03304}{{\ttfamily arXiv:1711.03304 [hep-ex]}}.

\bibitem{Sivers:1989cc}
D.~W. Sivers
\href{http://dx.doi.org/10.1103/PhysRevD.41.83}{{\em Phys. Rev.} {\bfseries
  D41} (1990) 83}.

\bibitem{Bravar:1999rq}
{\bfseries Spin Muon} Collaboration, A.~Bravar
  \href{http://dx.doi.org/10.1016/S0920-5632(99)00770-7}{{\em Nucl. Phys. Proc.
  Suppl.} {\bfseries 79} (1999) 520--522}.
[,520(1999)].

\bibitem{Airapetian:2001eg}
{\bfseries HERMES} Collaboration, A.~Airapetian {\em et~al.}
  \href{http://dx.doi.org/10.1103/PhysRevD.64.097101}{{\em Phys. Rev.}
  {\bfseries D64} (2001) 097101},
\href{http://arxiv.org/abs/hep-ex/0104005}{{\ttfamily arXiv:hep-ex/0104005
  [hep-ex]}}.

\bibitem{Ji:2013dva}
X.~Ji \href{http://dx.doi.org/10.1103/PhysRevLett.110.262002}{{\em Phys. Rev.
  Lett.} {\bfseries 110} (2013) 262002},
\href{http://arxiv.org/abs/1305.1539}{{\ttfamily arXiv:1305.1539 [hep-ph]}}.

\bibitem{Brodsky:2002cx}
S.~J. Brodsky, D.~S. Hwang, and I.~Schmidt
  \href{http://dx.doi.org/10.1016/S0370-2693(02)01320-5}{{\em Phys. Lett.}
  {\bfseries B530} (2002) 99--107},
\href{http://arxiv.org/abs/hep-ph/0201296}{{\ttfamily arXiv:hep-ph/0201296
  [hep-ph]}}.

\bibitem{Boer:2002ju}
D.~Boer, S.~J. Brodsky, and D.~S. Hwang
  \href{http://dx.doi.org/10.1103/PhysRevD.67.054003}{{\em Phys. Rev.}
  {\bfseries D67} (2003) 054003},
\href{http://arxiv.org/abs/hep-ph/0211110}{{\ttfamily arXiv:hep-ph/0211110
  [hep-ph]}}.

\bibitem{Bacchetta:2003rz}
A.~Bacchetta, A.~Schaefer, and J.-J. Yang
  \href{http://dx.doi.org/10.1016/j.physletb.2003.10.045}{{\em Phys. Lett.}
  {\bfseries B578} (2004) 109--118},
\href{http://arxiv.org/abs/hep-ph/0309246}{{\ttfamily arXiv:hep-ph/0309246
  [hep-ph]}}.

\bibitem{Lu:2004au}
Z.~Lu and B.-Q. Ma
  \href{http://dx.doi.org/10.1016/j.nuclphysa.2004.06.006}{{\em Nucl. Phys.}
  {\bfseries A741} (2004) 200--214},
\href{http://arxiv.org/abs/hep-ph/0406171}{{\ttfamily arXiv:hep-ph/0406171
  [hep-ph]}}.

\bibitem{Goeke:2006ef}
K.~Goeke, S.~Meissner, A.~Metz, and M.~Schlegel
  \href{http://dx.doi.org/10.1016/j.physletb.2006.05.004}{{\em Phys. Lett.}
  {\bfseries B637} (2006) 241--244},
\href{http://arxiv.org/abs/hep-ph/0601133}{{\ttfamily arXiv:hep-ph/0601133
  [hep-ph]}}.

\bibitem{Lu:2006kt}
Z.~Lu and I.~Schmidt \href{http://dx.doi.org/10.1103/PhysRevD.75.073008}{{\em
  Phys. Rev.} {\bfseries D75} (2007) 073008},
\href{http://arxiv.org/abs/hep-ph/0611158}{{\ttfamily arXiv:hep-ph/0611158
  [hep-ph]}}.

\bibitem{Bacchetta:2008af}
A.~Bacchetta, F.~Conti, and M.~Radici
  \href{http://dx.doi.org/10.1103/PhysRevD.78.074010}{{\em Phys. Rev.}
  {\bfseries D78} (2008) 074010},
\href{http://arxiv.org/abs/0807.0323}{{\ttfamily arXiv:0807.0323 [hep-ph]}}.

\bibitem{Maji:2017wwd}
T.~Maji, D.~Chakrabarti, and A.~Mukherjee
  \href{http://dx.doi.org/10.1103/PhysRevD.97.014016}{{\em Phys. Rev.}
  {\bfseries D97} no.~1, (2018) 014016},
\href{http://arxiv.org/abs/1711.02930}{{\ttfamily arXiv:1711.02930 [hep-ph]}}.

\bibitem{Yuan:2003wk}
F.~Yuan \href{http://dx.doi.org/10.1016/j.physletb.2003.09.052}{{\em Phys.
  Lett.} {\bfseries B575} (2003) 45--54},
\href{http://arxiv.org/abs/hep-ph/0308157}{{\ttfamily arXiv:hep-ph/0308157
  [hep-ph]}}.

\bibitem{Courtoy:2008mj}
A.~Courtoy, S.~Scopetta, and V.~Vento
  \href{http://dx.doi.org/10.1063/1.3122171}{{\em AIP Conf. Proc.} {\bfseries
  1105} no.~1, (2009) 193--196},
\href{http://arxiv.org/abs/0811.2368}{{\ttfamily arXiv:0811.2368 [hep-ph]}}.

\bibitem{Pasquini:2010af}
B.~Pasquini and F.~Yuan
  \href{http://dx.doi.org/10.1103/PhysRevD.81.114013}{{\em Phys. Rev.}
  {\bfseries D81} (2010) 114013},
\href{http://arxiv.org/abs/1001.5398}{{\ttfamily arXiv:1001.5398 [hep-ph]}}.

\bibitem{Belitsky:2002sm}
A.~V. Belitsky, X.~Ji, and F.~Yuan
  \href{http://dx.doi.org/10.1016/S0550-3213(03)00121-4}{{\em Nucl. Phys.}
  {\bfseries B656} (2003) 165--198},
\href{http://arxiv.org/abs/hep-ph/0208038}{{\ttfamily arXiv:hep-ph/0208038
  [hep-ph]}}.

\bibitem{Collins:2002kn}
J.~C. Collins \href{http://dx.doi.org/10.1016/S0370-2693(02)01819-1}{{\em Phys.
  Lett.} {\bfseries B536} (2002) 43--48},
\href{http://arxiv.org/abs/hep-ph/0204004}{{\ttfamily arXiv:hep-ph/0204004
  [hep-ph]}}.

\bibitem{Sullivan:1971kd}
J.~D. Sullivan
\href{http://dx.doi.org/10.1103/PhysRevD.5.1732}{{\em Phys. Rev.} {\bfseries
  D5} (1972) 1732--1737}.

\bibitem{Burkardt:2012hk}
M.~Burkardt, K.~S. Hendricks, C.-R. Ji, W.~Melnitchouk, and A.~W. Thomas
  \href{http://dx.doi.org/10.1103/PhysRevD.87.056009}{{\em Phys. Rev.}
  {\bfseries D87} no.~5, (2013) 056009},
\href{http://arxiv.org/abs/1211.5853}{{\ttfamily arXiv:1211.5853 [hep-ph]}}.

\bibitem{Salamu:2018cny}
Y.~Salamu, C.-R. Ji, W.~Melnitchouk, A.~W. Thomas, and P.~Wang
  \href{http://dx.doi.org/10.1103/PhysRevD.99.014041}{{\em Phys. Rev.}
  {\bfseries D99} no.~1, (2019) 014041},
\href{http://arxiv.org/abs/1806.07551}{{\ttfamily arXiv:1806.07551 [hep-ph]}}.

\bibitem{Salamu:2014pka}
Y.~Salamu, C.-R. Ji, W.~Melnitchouk, and P.~Wang
  \href{http://dx.doi.org/10.1103/PhysRevLett.114.122001}{{\em Phys. Rev.
  Lett.} {\bfseries 114} (2015) 122001},
\href{http://arxiv.org/abs/1409.5885}{{\ttfamily arXiv:1409.5885 [hep-ph]}}.

\bibitem{Wang:2016eoq}
X.~G. Wang, C.-R. Ji, W.~Melnitchouk, Y.~Salamu, A.~W. Thomas, and P.~Wang
  \href{http://dx.doi.org/10.1016/j.physletb.2016.09.014}{{\em Phys. Lett.}
  {\bfseries B762} (2016) 52--56},
\href{http://arxiv.org/abs/1602.06646}{{\ttfamily arXiv:1602.06646 [nucl-th]}}.

\bibitem{Adamczyk:2015gyk}
{\bfseries STAR} Collaboration, L.~Adamczyk {\em et~al.}
  \href{http://dx.doi.org/10.1103/PhysRevLett.116.132301}{{\em Phys. Rev.
  Lett.} {\bfseries 116} no.~13, (2016) 132301},
\href{http://arxiv.org/abs/1511.06003}{{\ttfamily arXiv:1511.06003 [nucl-ex]}}.

\bibitem{Huang:2015vpy}
J.~Huang, Z.-B. Kang, I.~Vitev, and H.~Xing
  \href{http://dx.doi.org/10.1103/PhysRevD.93.014036}{{\em Phys. Rev.}
  {\bfseries D93} no.~1, (2016) 014036},
\href{http://arxiv.org/abs/1511.06764}{{\ttfamily arXiv:1511.06764 [hep-ph]}}.

\bibitem{Bacchetta:2011gx}
A.~Bacchetta and M.~Radici
  \href{http://dx.doi.org/10.1103/PhysRevLett.107.212001}{{\em Phys. Rev.
  Lett.} {\bfseries 107} (2011) 212001},
\href{http://arxiv.org/abs/1107.5755}{{\ttfamily arXiv:1107.5755 [hep-ph]}}.

\bibitem{Anselmino:2016uie}
M.~Anselmino, M.~Boglione, U.~D'Alesio, F.~Murgia, and A.~Prokudin
  \href{http://dx.doi.org/10.1007/JHEP04(2017)046}{{\em JHEP} {\bfseries 04}
  (2017) 046},
\href{http://arxiv.org/abs/1612.06413}{{\ttfamily arXiv:1612.06413 [hep-ph]}}.

\bibitem{Martin:2017yms}
A.~Martin, F.~Bradamante, and V.~Barone
  \href{http://dx.doi.org/10.1103/PhysRevD.95.094024}{{\em Phys. Rev.}
  {\bfseries D95} no.~9, (2017) 094024},
\href{http://arxiv.org/abs/1701.08283}{{\ttfamily arXiv:1701.08283 [hep-ph]}}.

\bibitem{Boglione:2018dqd}
M.~Boglione, U.~D'Alesio, C.~Flore, and J.~O. Gonzalez-Hernandez
  \href{http://dx.doi.org/10.1007/JHEP07(2018)148}{{\em JHEP} {\bfseries 07}
  (2018) 148},
\href{http://arxiv.org/abs/1806.10645}{{\ttfamily arXiv:1806.10645 [hep-ph]}}.

\bibitem{Bacchetta:2004jz}
A.~Bacchetta, U.~D'Alesio, M.~Diehl, and C.~A. Miller
  \href{http://dx.doi.org/10.1103/PhysRevD.70.117504}{{\em Phys. Rev.}
  {\bfseries D70} (2004) 117504},
\href{http://arxiv.org/abs/hep-ph/0410050}{{\ttfamily arXiv:hep-ph/0410050
  [hep-ph]}}.

\bibitem{Ji:2002aa}
X.-d. Ji and F.~Yuan
  \href{http://dx.doi.org/10.1016/S0370-2693(02)02384-5}{{\em Phys. Lett.}
  {\bfseries B543} (2002) 66--72},
\href{http://arxiv.org/abs/hep-ph/0206057}{{\ttfamily arXiv:hep-ph/0206057
  [hep-ph]}}.

\bibitem{Chen:2001nb}
J.-W. Chen and X.-d. Ji \href{http://dx.doi.org/10.1103/PhysRevLett.87.152002,
  10.1103/PhysRevLett.88.249901}{{\em Phys. Rev. Lett.} {\bfseries 87} (2001)
  152002}, \href{http://arxiv.org/abs/hep-ph/0107158}{{\ttfamily
  arXiv:hep-ph/0107158 [hep-ph]}}.
[Erratum: Phys. Rev. Lett.88,249901(2002)].

\bibitem{Bando:1984ej}
M.~Bando, T.~Kugo, S.~Uehara, K.~Yamawaki, and T.~Yanagida
\href{http://dx.doi.org/10.1103/PhysRevLett.54.1215}{{\em Phys. Rev. Lett.}
  {\bfseries 54} (1985) 1215}.

\bibitem{Bando:1987br}
M.~Bando, T.~Kugo, and K.~Yamawaki
\href{http://dx.doi.org/10.1016/0370-1573(88)90019-1}{{\em Phys. Rept.}
  {\bfseries 164} (1988) 217--314}.

\bibitem{Tanabashi:1995nz}
M.~Tanabashi \href{http://dx.doi.org/10.1016/0370-2693(96)00827-1}{{\em Phys.
  Lett.} {\bfseries B384} (1996) 218--226},
\href{http://arxiv.org/abs/hep-ph/9511367}{{\ttfamily arXiv:hep-ph/9511367
  [hep-ph]}}.

\bibitem{Kawarabayashi:1966kd}
K.~Kawarabayashi and M.~Suzuki
\href{http://dx.doi.org/10.1103/PhysRevLett.16.255}{{\em Phys. Rev. Lett.}
  {\bfseries 16} (1966) 255}.

\bibitem{Riazuddin:1966sw}
Riazuddin and Fayyazuddin
\href{http://dx.doi.org/10.1103/PhysRev.147.1071}{{\em Phys. Rev.} {\bfseries
  147} (1966) 1071--1073}.

\bibitem{Tanabashi:2018oca}
{\bfseries Particle Data Group} Collaboration, M.~Tanabashi {\em et~al.}
\href{http://dx.doi.org/10.1103/PhysRevD.98.030001}{{\em Phys. Rev.} {\bfseries
  D98} no.~3, (2018) 030001}.

\bibitem{Scherer:2002tk}
S.~Scherer {\em Adv. Nucl. Phys.} {\bfseries 27} (2003) 277,
\href{http://arxiv.org/abs/hep-ph/0210398}{{\ttfamily arXiv:hep-ph/0210398
  [hep-ph]}}.

\bibitem{Schechter:1986vs}
J.~Schechter
\href{http://dx.doi.org/10.1103/PhysRevD.34.868}{{\em Phys. Rev.} {\bfseries
  D34} (1986) 868}.

\bibitem{Yamawaki:1986zz}
K.~Yamawaki
\href{http://dx.doi.org/10.1103/PhysRevD.35.412}{{\em Phys. Rev.} {\bfseries
  D35} (1987) 412}.

\bibitem{Meissner:1986tc}
U.~G. Meissner and I.~Zahed
{\em Z. Phys.} {\bfseries A327} (1987) 5--15.

\bibitem{Jenkins:1991ts}
E.~E. Jenkins
\href{http://dx.doi.org/10.1016/0550-3213(92)90203-N}{{\em Nucl. Phys.}
  {\bfseries B368} (1992) 190--203}.

\bibitem{Geng:2009ys}
L.~S. Geng, J.~Martin~Camalich, and M.~J. Vicente~Vacas
  \href{http://dx.doi.org/10.1103/PhysRevD.80.034027}{{\em Phys. Rev.}
  {\bfseries D80} (2009) 034027},
\href{http://arxiv.org/abs/0907.0631}{{\ttfamily arXiv:0907.0631 [hep-ph]}}.

\bibitem{Jones:1972ky}
H.~F. Jones and M.~D. Scadron
\href{http://dx.doi.org/10.1016/0003-4916(73)90476-4}{{\em Annals Phys.}
  {\bfseries 81} (1973) 1--14}.

\bibitem{Brown:1975di}
G.~E. Brown and W.~Weise
\href{http://dx.doi.org/10.1016/0370-1573(75)90026-5}{{\em Phys. Rept.}
  {\bfseries 22} (1975) 279--337}.

\bibitem{Mergell:1995bf}
P.~Mergell, U.~G. Meissner, and D.~Drechsel
  \href{http://dx.doi.org/10.1016/0375-9474(95)00339-8}{{\em Nucl. Phys.}
  {\bfseries A596} (1996) 367--396},
\href{http://arxiv.org/abs/hep-ph/9506375}{{\ttfamily arXiv:hep-ph/9506375
  [hep-ph]}}.

\bibitem{Haidenbauer:2017sws}
J.~Haidenbauer, S.~Petschauer, N.~Kaiser, U.-G. Mei{\ss}ner, and W.~Weise
  \href{http://dx.doi.org/10.1140/epjc/s10052-017-5309-4}{{\em Eur. Phys. J.}
  {\bfseries C77} no.~11, (2017) 760},
\href{http://arxiv.org/abs/1708.08071}{{\ttfamily arXiv:1708.08071 [nucl-th]}}.

\bibitem{Aicher:2010cb}
M.~Aicher, A.~Schafer, and W.~Vogelsang
  \href{http://dx.doi.org/10.1103/PhysRevLett.105.252003}{{\em Phys. Rev.
  Lett.} {\bfseries 105} (2010) 252003},
\href{http://arxiv.org/abs/1009.2481}{{\ttfamily arXiv:1009.2481 [hep-ph]}}.

\bibitem{Machleidt:1989tm}
R.~Machleidt
{\em Adv. Nucl. Phys.} {\bfseries 19} (1989) 189--376.

\bibitem{Barry:2018ort}
P.~C. Barry, N.~Sato, W.~Melnitchouk, and C.-R. Ji
  \href{http://dx.doi.org/10.1103/PhysRevLett.121.152001}{{\em Phys. Rev.
  Lett.} {\bfseries 121} no.~15, (2018) 152001},
\href{http://arxiv.org/abs/1804.01965}{{\ttfamily arXiv:1804.01965 [hep-ph]}}.

\bibitem{He:2018eyz}
F.~He and P.~Wang \href{http://dx.doi.org/10.1103/PhysRevD.98.036007}{{\em
  Phys. Rev.} {\bfseries D98} no.~3, (2018) 036007},
\href{http://arxiv.org/abs/1805.11986}{{\ttfamily arXiv:1805.11986 [hep-ph]}}.

\bibitem{Yu}
Y.~Salamu, C.-R. Ji, W.~Melnitchouk, A.~W. Thomas, P.~Wang and X.~G.~Wang
\href{http://arxiv.org/abs/1907.08551}{{\ttfamily arXiv:1907.08551 [hep-ph]}}.

\bibitem{Pobylitsa:2003ty}
P.~V. Pobylitsa
\href{http://arxiv.org/abs/hep-ph/0301236}{{\ttfamily arXiv:hep-ph/0301236
  [hep-ph]}}.

\bibitem{Aybat:2011ge}
S.~M. Aybat, J.~C. Collins, J.-W. Qiu, and T.~C. Rogers
  \href{http://dx.doi.org/10.1103/PhysRevD.85.034043}{{\em Phys. Rev.}
  {\bfseries D85} (2012) 034043},
\href{http://arxiv.org/abs/1110.6428}{{\ttfamily arXiv:1110.6428 [hep-ph]}}.

\bibitem{Anselmino:2012aa}
M.~Anselmino, M.~Boglione, and S.~Melis
  \href{http://dx.doi.org/10.1103/PhysRevD.86.014028}{{\em Phys. Rev.}
  {\bfseries D86} (2012) 014028},
\href{http://arxiv.org/abs/1204.1239}{{\ttfamily arXiv:1204.1239 [hep-ph]}}.

\bibitem{Kang:2012em}
Z.-B. Kang and J.-W. Qiu
  \href{http://dx.doi.org/10.1016/j.physletb.2012.06.021}{{\em Phys. Lett.}
  {\bfseries B713} (2012) 273--276},
\href{http://arxiv.org/abs/1205.1019}{{\ttfamily arXiv:1205.1019 [hep-ph]}}.

\end{thebibliography}
\providecommand{\href}[2]{#2}\begingroup\raggedright\endgroup

\end{document}